\documentclass[%
 aip,
 amsmath,amssymb,
 reprint,%
]{revtex4-1}

\usepackage{graphicx}
\usepackage{dcolumn}
\usepackage{bm}

\usepackage[utf8]{inputenc}
\usepackage[T1]{fontenc}
\usepackage{mathptmx}

\usepackage{units}
\usepackage{mathtools}
\DeclareMathOperator{\arctanh}{arctanh}
\usepackage{hyperref}
\usepackage{color}
\usepackage{lineno}

\begin{document}

\preprint{AIP/123-QED}

\title[All-Optical and Microwave-Free Detection of Meissner Screening using Nitrogen Vacancy Centers in Diamond]{All-Optical and Microwave-Free Detection of Meissner Screening using Nitrogen-Vacancy Centers in Diamond}

\author{D. Paone}
\affiliation{%
	Max Planck Institute for Solid State Research
}%
\affiliation{%
	3$^{rd}$ Institute of Physics and Institute for Integrated Quantum Science and Technology IQST, University Stuttgart
}%

\author{D. Pinto}%
\affiliation{%
	Max Planck Institute for Solid State Research
}%
\affiliation{%
	Institut de Physique, \'{E}cole Polytechnique F\'{e}d\'{e}rale de Lausanne
}%

\author{G. Kim}
\affiliation{%
	Max Planck Institute for Solid State Research
}%

\author{L. Feng}
\affiliation{%
	Max Planck Institute for Solid State Research
}%
\affiliation{4$^{th}$ Institute of Physics and Research Center SCoPE, University Stuttgart}

\author{M-J. Kim}
\affiliation{%
	Max Planck Institute for Solid State Research
}%
\affiliation{4$^{th}$ Institute of Physics and Research Center SCoPE, University Stuttgart}

\author{R. Stöhr}
\affiliation{%
	3$^{rd}$ Institute of Physics and Institute for Integrated Quantum Science and Technology IQST, University Stuttgart
}%

\author{A. Singha}
\affiliation{%
	Max Planck Institute for Solid State Research
}%

\author{S. Kaiser}
\affiliation{%
	Max Planck Institute for Solid State Research
}%
\affiliation{4$^{th}$ Institute of Physics and Research Center SCoPE, University Stuttgart}

\author{G. Logvenov}
\affiliation{%
	Max Planck Institute for Solid State Research
}%

\author{B. Keimer}
\affiliation{%
	Max Planck Institute for Solid State Research
}%

\author{J. Wrachtrup}
\affiliation{%
	Max Planck Institute for Solid State Research
}%
\affiliation{%
	3$^{rd}$ Institute of Physics and Institute for Integrated Quantum Science and Technology IQST, University Stuttgart
}%

\author{K. Kern}
\affiliation{%
	Max Planck Institute for Solid State Research
}%
\affiliation{%
	Institut de Physique, \'{E}cole Polytechnique F\'{e}d\'{e}rale de Lausanne
}%

\date{\today}

\begin{abstract}
Microscopic studies on thin film superconductors play an important role for probing non-equilibrium phase transitions and revealing dynamics at the nanoscale.
However, magnetic sensors with nanometer scale spatial and picosecond temporal resolution are essential for exploring these.	
Here, we present an all-optical, microwave-free method, that utilizes the negatively charged nitrogen-vacancy (NV) center in diamond as a non-invasive quantum sensor and enables the spatial detection of the Meissner state in a superconducting thin film. We place an NV implanted diamond membrane on a $\unit[20]{nm}$ thick superconducting La$_{2-x}$Sr$_x$CuO$_4$ (LSCO) thin film with $T_c$ of $\unit[34]{K}$. The strong B-field dependence of the NV photoluminescence (PL) allows us to investigate the Meissner screening in LSCO under an externally applied magnetic field of $\unit[4.2]{mT}$ in a non-resonant manner. The magnetic field profile along the LSCO thin film can be reproduced using Brandt's analytical model, revealing a critical current density $j_c$ of $\unit[1.4 \cdot 10^8]{A/cm^2}$. Our work can be potentially extended further with a combination of optical pump probe spectroscopy, for the local detection of time-resolved dynamical phenomena in nanomagnetic materials. 
\end{abstract}

\maketitle

%
\section{Introduction}
Microscopic phenomena revealing complex magnetic phases in two-dimensional materials are catching the central attention in modern condensed-matter physics \cite{Bandy}. Prime examples are superconducting systems which are accompanied by electronic phases \cite{Wu, Bednorz, Hazen}, such as vortex formation in the case of type II superconductors \cite{Essmann}. Various approaches are already established for studying superconductivity mainly based on superconducting quantum interference devices (SQUIDs) \cite{Ceccarelli}, magnetic force microscopy (MFM) \cite{Hug}, scanning tunneling microscopy (STM) \cite{Kot} and the investigation of magneto-optical effects \cite{Goa}. However, each of these techniques suffer from drawbacks such as limited temperature and magnetic field ranges, spatial resolution and complex sample preparation. A promising alternative for surpassing these drawbacks is to employ negatively charged nitrogen vacancy (NV) centers in diamond. In fact, the NV center in diamond is a non-invasive nanoscale magnetic field sensor allowing measurements at both, cryogenic as well as ambient conditions, with a magnetic field sensitivity of $\unit[1]{pT/\sqrt{Hz}}$ for NV ensembles and $\unit[1]{\mu T/\sqrt{Hz}}$ in the case of single NV centers \cite{Degen}. Applications of NV center sensing have been shown in single molecular systems, investigated using nuclear magnetic resonance (NMR) \cite{Staudacher} and electron spin resonance (ESR) \cite{Schlipf, Pinto}. In addition, magnetic properties of materials have been investigated including spin waves \cite{Du}, ferromagnetism \cite{Page, Ferror} and superconductivity \cite{Waxman, Thiel, Pelliccione, Nusran, Joshi, Xu, Rohner, Sydney, SCreview} at the nanoscale with this approach. The fundamental sensing principle of the NV center relies on the spin dependent photoluminescence (PL) of the defect center \cite{Wrachtrup, Donghun}. Microwave excitations allow coherent manipulation within different spin sublevels present in the ground state. The resulting transition frequencies show a Zeeman effect, forming a toolset for magnetic field sensing, known as optically detected magnetic resonance (ODMR) technique. 
However, microwave excitations are usually linked with heating effects, which could locally change properties of the investigated sample \cite{Sydney}.\\
Here, we introduce the magnetic field dependent fluorescence yield of an NV center ensemble which allows a reliable and direct investigation of the spatial modulation of superconductivity on a microscopic scale. Besides calibrating the change in the NV emission intensity caused by an external magnetic field with ODMR spectroscopy, our experiments rely solely on an all-optical, non-resonant, microwave-free measurement scheme. We position an NV implanted diamond membrane at the edge of a superconducting La$_{2-x}$Sr$_x$CuO$_4$ (LSCO) thin film. We collect the NV center fluorescence at different positions on the superconductor at $\unit[4.2]{K}$ and under an external magnetic field of $\unit[4.2]{mT}$ aligned along the $z$-direction. Normalization to zero-field measurements provides us a fingerprint of the Meissner screening in terms of the PL rate drop when the NV ensembles are in close proximity ($\approx \unit[1]{\mu m}$) to the LSCO thin film. Combining this with an analytical model, developed by E. H. Brandt \cite{Brandt}, allows us to extract the critical current density $j_c$. Furthermore, we provide a comparison of our results with a complementary SQUID measurement.
%
\begin{figure}
	\centering
	\includegraphics[scale=1.0]{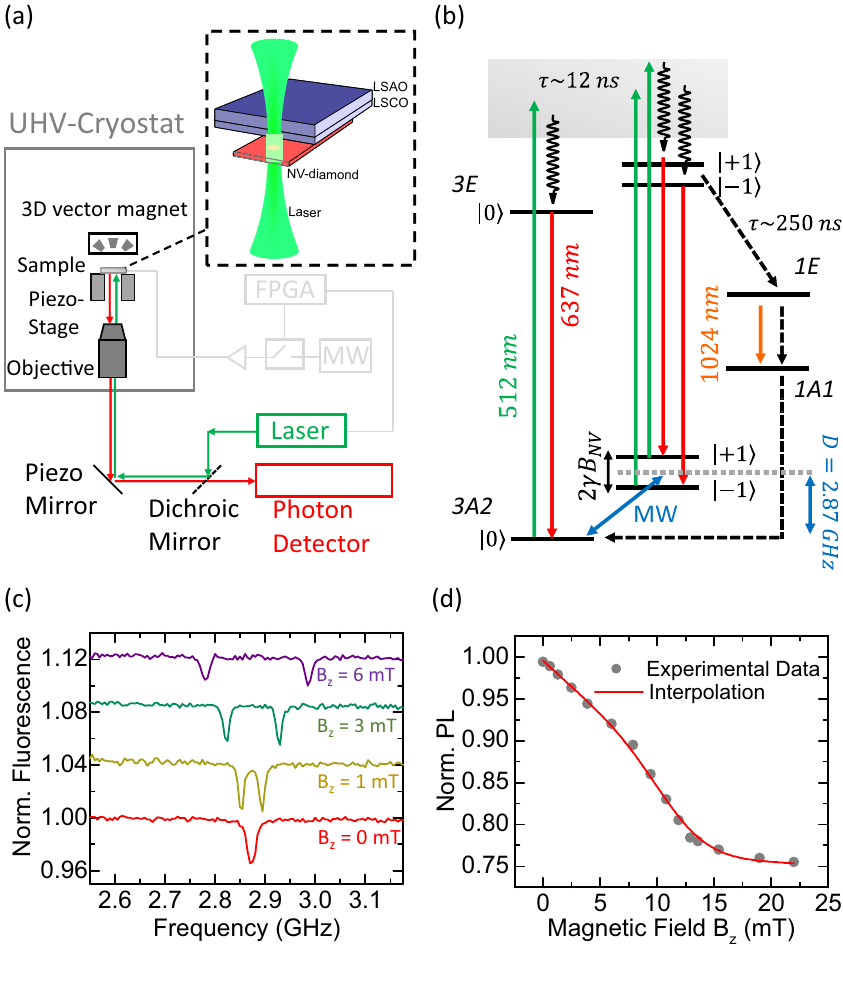}
	\caption{(a) Sketch of the experimental setup. A confocal microscopy setup is attached to a UHV-cryostat at $\unit[4.2]{K}$ and equipped with a 3D vector magnet for applying external magnetic fields. The microwave circuit is only used for resonant calibration purpose and is shown in light grey. The inset depicts the sample geometry consisting of an NV diamond membrane attached onto a superconducting LSCO thin film. Dimensions are not drawn to scale. (b) Energy level scheme of the negatively charged NV center. The ground-state spin levels can be pumped via a green laser to the excited state. The relaxation to the ground-state leads to an emission of red photons, which are measured using a confocal microscope. The NV center spin state can be driven between the $m_s=|0\rangle$ and the $m_s=|-1\rangle$, $m_s=|+1\rangle$ states using microwave excitations. The magnetic field experienced by the NVs can be calculated using the Zeeman equation. (c) ODMR spectra of an NV ensemble in the bare diamond membrane for different magnetic fields aligned in z-direction, acquired at $\unit[4.2]{K}$ inside the UHV cryostat. The spectra are vertically offset for clarity. (d) Effect of increasing the external magnetic field on the observed NV fluorescence. The count rate drop is collected for different magnetic fields aligned in z-direction, estimated from the corresponding ODMR measurements. The red solid line indicates a smoothing spline interpolation. The measurement has been performed at ambient conditions in a separate confocal setup with a permanent magnet attached closely to the diamond membrane.}
	\label{abb: main1}
\end{figure}
\section{Experimental Setup and Methods}
All measurements are carried out using a confocal microscope connected to an UHV-He bath cryostat operating at a base pressure of $\unit[3 \cdot 10^{-10}]{mbar}$ at $\unit[4.2]{K}$ (Fig.~\ref{abb: main1}(a)) \cite{Nolte}. A green $\unit[512]{nm}$ pulsed laser is used to excite the NV centers. The emitted fluorescence is recorded with a photon detector device. The laser spot is scanned over the sample, while recording the NV fluorescence, resulting in a confocal image. Further details of this experimental setup and of an additional setup used for NV characterization at ambient conditions can be found in section 1 of the Supporting Information (SI). The pulsed microwave source integrated with the setup allows us to perform magnetometry with NV centers and ODMR measurements which require NV spin manipulations.\\
Magnetometry with NV centers relies on the energy-level scheme shown in Fig.~\ref{abb: main1}(b). The ground electronic state of the NV center is a spin triplet. The energy gap between the excited triplet state and the ground state corresponds to a photon emission of $\unit[637]{nm}$. The NV center can be excited from the ground state into the phonon side band using $\approx \unit[512]{nm}$ green laser light. Subsequently, it relaxes to the ground state by emitting photons in the range of 637-$\unit[750]{nm}$. The fluorescence is highly spin state selective \cite{Wrachtrup}. In particular, the fluorescence rate of the $m_s=|-1\rangle,|+1\rangle$ states is lower than that of the $m_s=|0\rangle$. In presence of an externally applied magnetic field, the $|\pm 1 \rangle$ states experience a Zeeman splitting. Subsequently, a resonant microwave excitation enables the transition between the bright $m_s=|0\rangle$ state and one of the less fluorescent dark $|\pm 1\rangle$ states, whenever the microwave frequency matches the induced energy splitting. Therefore, in absence of an external magnetic field, the NV fluorescence shows only a single resonance signifying the zero field splitting of $D = \unit[2.87]{GHz}$. For a non zero magnetic field, this splits into two resonances corresponding to the $|0\rangle \rightarrow |-1\rangle$ and $|0\rangle \rightarrow |+1\rangle$ transitions. The unique combination of these properties of the NV center, allows us to calibrate the magnetic field, by observing field dependent ODMR spectra, as shown in Fig.~\ref{abb: main1}(c). For all measurements, the magnetic field is applied along the z direction ($\vec{B} = (0, 0, B_z)$), normal to the (100) surface. Therefore, all four possible NV orientations experience the same Zeeman splitting $\Delta f$ \cite{Nusran} given by $\Delta f = \frac{2 \gamma B_{z}}{\sqrt{3}}$. 
$\gamma$ is the gyromagnetic ratio, which is $\unit[28]{MHz/mT}$ for the NV electronic spin. 
The observed frequency splitting is proportional to the applied magnetic field following this equation.
However, this method requires the application of a resonant microwave driving frequency. Microwave applications are often accompanied by local heating effects, which could cause undesired changes in the properties of the investigated system, especially in case of a superconducting sample.\\
This can be circumvented by utilizing a non-resonant measurement scheme without microwave excitations. The fluorescence yield detected from the NV center strongly depends on the applied off-axis magnetic field. In particular, it decreases with increasing the $B_z$-field. This can be explained by the spin mixing of the sublevels when the magnetic field is misaligned with respect to the NV axis. The mixing of the spin sublevels leads to an inefficient spin-dependent PL rate by enhancing the probablity of the non-radiative inter system crossing to the metastable state. The resulting PL drop can be used for a qualitative investigation of the magnetic fields \cite{Tetienne}.
Fig.~\ref{abb: main1}(d) shows this behavior of the NV emission for different magnetic fields $B_z$ at ambient conditions. A significant decrease of the detected PL up to $\unit[25]{\%}$ is observed for the highest applied magnetic field of $\unit[22.5]{mT}$ along the $z$-direction. This approach still relies on ODMR spectra for the quantitative calibration of the magnetic fields on the NV axis. However, besides this the direct all-optical record of the NV emission relies on a non-resonant, microwave-free method. Measurements of field variations and qualitative observations of magnetic properties, which are often required for supercondcting systems, are still viable using the direct emission of the NV centers. Therefore, the magnetic field dependent PL can be a sensitive tool for detecting non-equilibrium phase transitions and dynamical phenomena with a high spatial resolution.
\begin{figure}
	\centering
	\includegraphics[scale=1.0]{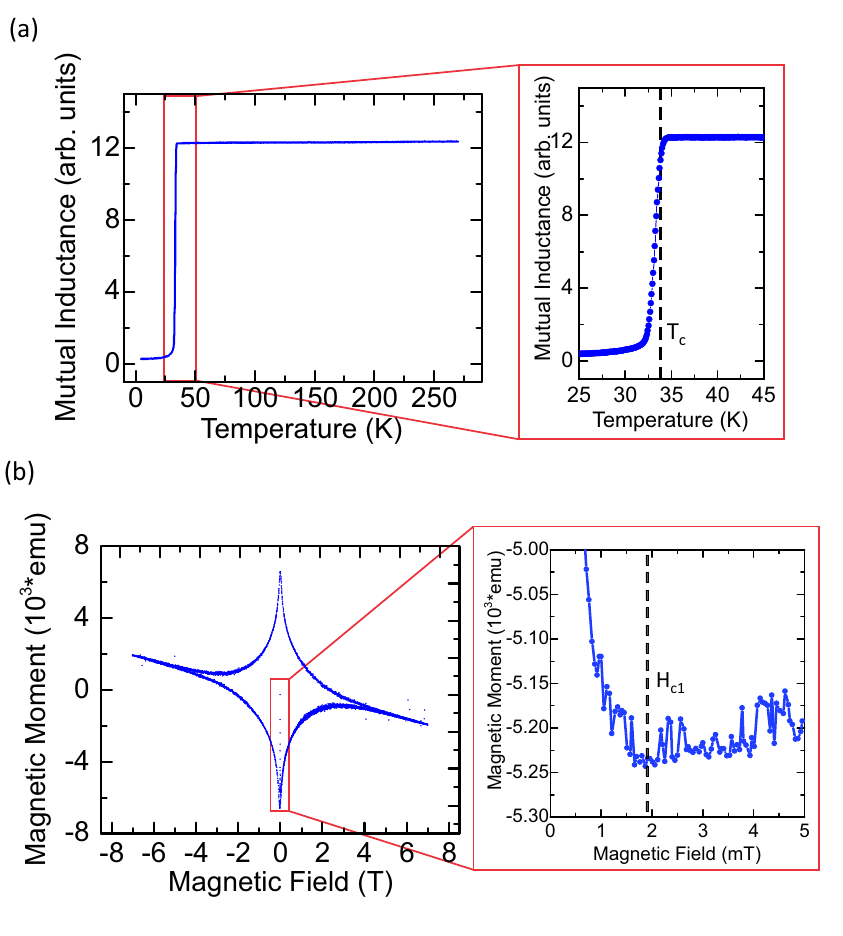}
	\caption{(a) Real part of the mutual inductance versus the temperature. The mutual inductance is measured as the magnetic interaction between two coils. Within these, the LSCO sample is positioned and the temperature is swept from $\unit[300]{K}$ to $\unit[5]{K}$. The inset shows a significant drop of the inductance at $\unit[34]{K}$ indicating a critical temperature of $T_c =\unit[34]{K}$. (b) Hysteresis curve measured with a superconducting quantum interference device (SQUID) at $\unit[4]{K}$. The magnetic moment is investigated as a function of the applied magnetic field ranging from $\unit[-7]{T}$ to $\unit[7]{T}$. The inset shows the behavior of the magnetic moment for weak magnetic fields between 0 and $\unit[5]{mT}$. The inflection point of the curve indicates the first critical field $H_{c1}$, being at around $\unit[2]{mT}$.}
	\label{abb: main2}
\end{figure}
\section{Results and Discussion}
\subsection{Investigated LSCO Sample}
In order to demonstrate this, we characterized the Meissner screening caused by a type II cuprate superconducting LSCO sample. LSCO is among the most studied high $T_c$ superconductors in recent years \cite{Logvenov, Gilardi}. Furthermore, it has attracted much interest since Cooper pair formation, diamagnetism \cite{Li} and vortex mechanisms \cite{Iguchi} have been measured above $T_c$. Therefore, LSCO is not only an ideal sample to benchmark our technique with existing quantities and models, but also forms a system with interesting superconducting properties which could be investigated with our method in future experiments. The studied sample consists of a single crystalline LSCO thin film epitaxially grown on a (001) LaSrAlO$_4$ (LSAO) substrate \cite{Gideok} and exhibits a critical temperature of $T_c=\unit[34]{K}$ which is depicted in the mutual inductance measurement shown in Fig~\ref{abb: main2}(a). The mutual inductance has been measured by placing the LSCO sample within two pick-up coils (see section 7 of the SI). By applying a weak magnetic field through one of the coils, the magnetic interaction through the superconducting sample can be investigated. We also recorded the magnetic moment $m$ as a function of the magnetic field at \unit[4]{K} for gaining more information about the magnetic phase diagram including the lower critical field $H_{c1}$ (see Fig.~\ref{abb: main2}(b)). This indicates a first magnetic penetration at about $\unit[2]{mT}$, where magnetic fluxes start to enter the superconducting sample. 
\begin{figure}
	\centering
	\includegraphics[scale=1.0]{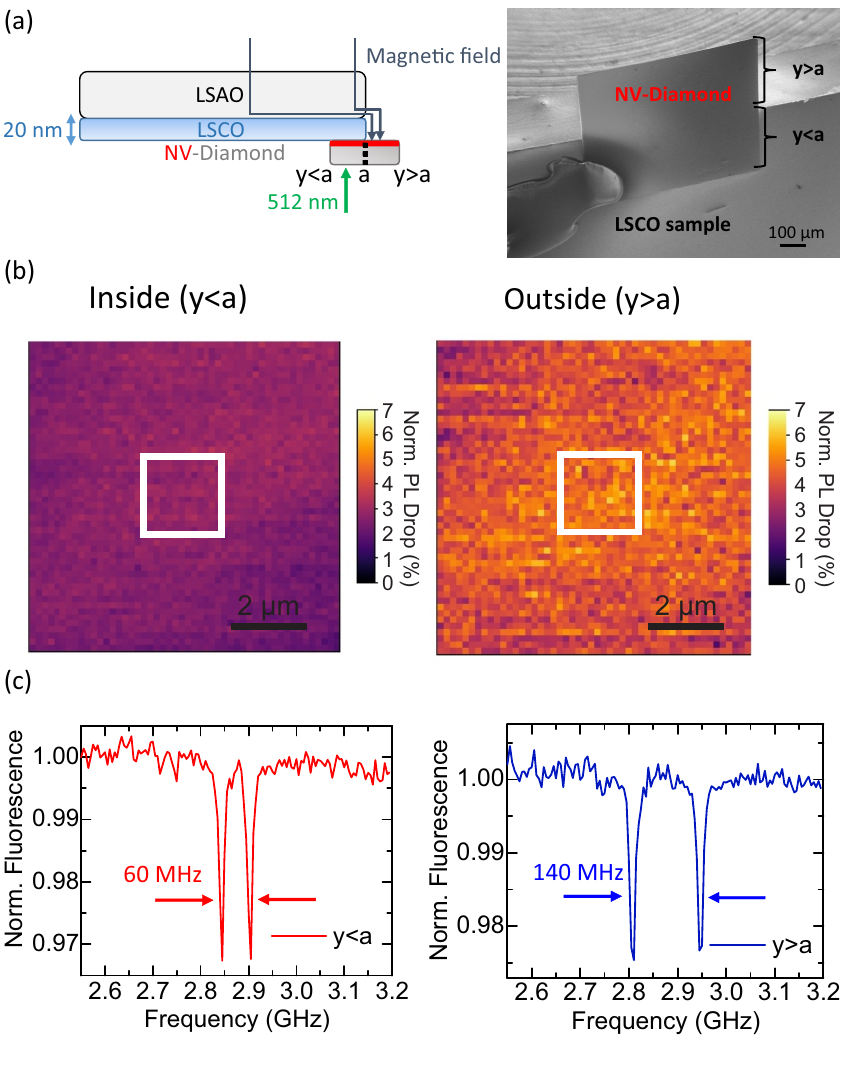}
	\caption{(a) Schematic and scanning electron microscope image (SEM) of the investigated system showing an NV implanted diamond membrane glued at the edge of a thin film superondcuting LSCO. Due to the Meissner effect, an applied magnetic field gets expelled by the LSCO thin film. An enhancement of the magnetic flux density occurs at the edges of the superconducting sample. The dimensions are not drawn to scale. (b) Normalized confocal images indicating the NV fluorescence behavior inside ($y<a$) and outside ($y>a$) of the superconductor at $\unit[4.2]{K}$. The normalized fluorescence count rate drop is determined by averaging the PL counts from the central $\unit[2]{\mu m}$ x $\unit[2]{\mu m}$ area, highlighted as white box. We observe almost a two-fold decrease in the PL drop inside the sample, indicating the Meissner screening of the LSCO thin film. All data are normalized with respect to images measured at $B=0$. The confocal scans are obtained with a resolution of $100$ x $100$ pixels, where each pixel has been recorded for $\unit[0.01]{s}$. (c) Effect of Meissner screening verified by using ODMR spectra for $y>a$ and $y<a$ at $\unit[4.2]{K}$. For $y>a$ (blue curve), the ODMR splitting is about 2 times larger compared to the case of $y<a$ (red curve).}
	\label{abb: main3}
\end{figure}
\subsection{Detection of Meissner State with the NV Center PL Drop} 
To study the spatial distribution of the Meissner screening caused by the LSCO thin film at $\unit[4.2]{K}$, we positioned our diamond membrane across the edge of the LSCO sample (see Fig.~\ref{abb: main3}(a)). This particular geometry allows us to characterize the effect of an external magnetic field on the LSCO thin film $(y<a)$, as well as background measurement without having LSCO on top $(y>a)$. Note that all PL measurements are normalized with respect to a corresponding zero field confocal scan for estimating the effective fluorescence rate drop. Fig.~\ref{abb: main3}(b) shows a confocal scan of the NV ensemble obtained from regions $y<a$ (left panel) and $y>a$ (right panel) respectively, for an applied field of $\unit[4.2]{mT}$. It is evident that the zero-field normalized confocal scan obtained from $y>a$ is significantly brighter compared to the case of $y<a$. We choose the central area (depicted as white square) of the confocal images for evaluation, since the center of the image shows the least distortion (see section 2 of the SI). For $y<a$, the normalization with respect to a zero field measurement reveals an average fluorescence drop of $\unit[2.5]{\%}$ in the central area. In contrast, for $y>a$ the fluorescence count rate drop amounts to $\unit[5.4]{\%}$. The $\unit[2.9]{\%}$ increase in the fluorescence count rate drop can be explained by the presence of the LSCO thin film. The superconducting phase in the LSCO thin film causes a Meissner expulsion \cite{Takeda} leading to varying magnetic fields experienced over the whole diamond membrane. NV centers in the area of $y<a$ experience a reduced magnetic field strength and therefore show a significantly decreased PL drop compared to the $y>a$ region. For $y>a$, which is outside the superconductor, no Meissner screening is present. Therefore, the NV center ensemble measures the unscreened applied field of $\unit[4.2]{mT}$ resulting in a higher fluorescence drop. Note that, the confocal scans have been obtained by a fixed laser power of $\unit[0.688]{\mu W}$, measured in front of the UHV chamber glass with a powermeter. With such low laserpower we can ensure, that the laser spot is not changing the superconducting properties of our sample \cite{Sydney}.\\ 
In order to quantify the strength of the magnetic field that penetrates through the LSCO thin film, we first verify the Meissner effect using a resonant process, i.e. ODMR spectroscopy. We obtained ODMR spectra on the NV ensembles for both regions, $y<a$ and $y>a$. The corresponding spectra are depicted in Fig.~\ref{abb: main3}(c). The red curve on the left panel shows the resonance peaks of the NV ensemble inside the LSCO thin film with a frequency splitting of $\Delta f = \unit[60]{MHz}$. This corresponds to a $z$-aligned magnetic field of $\approx \unit[1.8]{mT}$. In contrast, the blue curve (Fig.~\ref{abb: main3}(c), right panel), which represents the ODMR spectrum of the NV ensemble for $y>a$, exhibits a splitting $\Delta f$ of $\approx \unit[140]{MHz}$ corresponding to a magnetic field strength of $\approx \unit[4.2]{mT}$ along the $z$-direction. This is in an excellent agreement with the fact, that the NV ensemble in $y>a$ does not experience any Meissner screening. Therefore, the magnetic field strength calculated from the ODMR spectrum matches the applied, unscreened magnetic field value. Consequently, these results corroborate a magnetic field screening of $\approx \unit[56]{\%}$. Note that, the magnetic field in the region of $y<a$ is not vanishing. This can be attributed to the relatively large NV to superconductor distance ($\approx \unit[1]{\mu m}$) (see section 6 of the SI).  Also the formation of magnetic vortices in type II superconductors could result in a non vanishing magnetic field in the LSCO sample \cite{Gilardi}.\\
\begin{figure}
	\centering
	\includegraphics[scale=1.0]{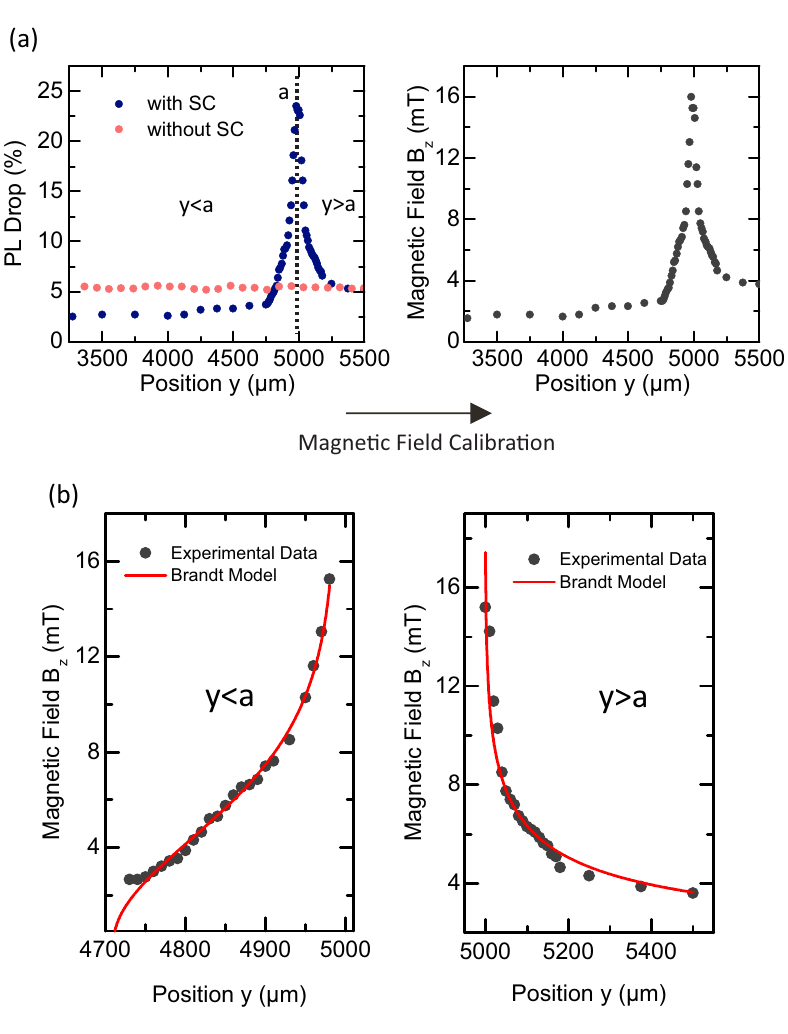}
	\caption{(a) Spatial variation in the PL drop measured by laser raster scan at $\unit[4.2]{K}$ with (blue dots) and without (red dots) superconducting sample (left panel). At each position, the PL drop of the NV ensemble is recorded. Estimates of the magnetic fields (right panel) come from a calibration with ODMR spectra. In presence of the LSCO thin film, for $y>a$, the external magnetic field asymptotically reaches $\unit[4.2]{mT}$, while for $y<a$ it is strongly screened due to the superconducting properties of LSCO. The sharp increase of the magnetic field at the edge indicates high magnetic flux densities. Such a spatial dependence is not observed in absence of the LSCO thin film. Each data point corresponds to the normalized averaged PL drop at the center of the confocal scan of about $\unit[4]{\mu m^2}$. (b) Fitting of the experimental data by using Brandt's model. The solid lines represent the fit. Both fit functions (inside and outside) reveal a critical current density $j_c$ of $\unit[1.4 \cdot 10^8]{A/cm^2}$.}
	\label{abb: main4}
\end{figure}
This approach can be extended further to gain insights about the Meissner screening and for characterizing the LSCO thin film to a greater detail. This is achieved by raster sweeping the laser focal spot over the diamond membrane along the $y$ direction. The corresponding spatial variation of the PL drop under an applied magnetic field of $B_z=\unit[4.2]{mT}$ is shown in Fig.~\ref{abb: main4}(a) (left panel). Again, each confocal scan and corresponding fluorescence rate is normalized to a zero field measurement for evaluating the relative fluorescence drop. Two distinct regimes characterized by different degrees of PL drops, are observed for $y<a$ and $y>a$ respectively. Inside the superconductor, we measure a relatively homogeneous PL drop of about $\unit[2]{\%}$, which slowly increases towards the boundary, i.e. $y=a$. Note that, the conversion from PL drop to an effective magnetic field is achieved using the calibration data presented in Fig.~\ref{abb: main1}(d). The latter reveals a magnetic field strength ranging between $\unit[1.8]{mT}$ and $\unit[2.1]{mT}$, as plotted on the right panel of Fig.~\ref{abb: main4}(a). In contrast, an increased PL drop of $\approx \unit[5.1]{\%}$ is observed at locations which are far away from the LSCO thin film. This corresponds to a magnetic field strength of $\approx \unit[4]{mT}$ which is in good agreement with the applied magnetic field. In addition, we have also observed a significant and sharp increase in the PL drop at the boundary, i.e., $y=a$, implying an equivalent increase of the effective magnetic flux density, which penetrates through the LSCO thin film. While the majority of the LSCO thin film expels the magnetic field, near the boundaries the field is enhanced to almost $\unit[16]{mT}$. The magnetic field flux is screened to the edge of the LSCO sample due to the diamagnetic properties of the superconducting compound. This leads to an enhanced magnetic flux density at the boundary of the superconducting LSCO sample. Note that, in absence of the LSCO sample, such a spatial dependence of the PL drop is not observed in the bare NV membrane (red dots in Fig.~\ref{abb: main4}(a)).
\subsection{Application of Brandt's Model}
The magnetic field profile in a superconducting thin film can be analytically evaluated by Brandt's model\cite{Brandt}. The spatial dependence along the $y$-direction, of the magnetic field applied perpendicular to the superconducting thin film is:			
\begin{align}
H(y) = \begin{cases}
\frac{J_c}{\pi} \arctanh \frac{\sqrt{ \left(y^2 - b^2 \right)}}{c |y|}  & y<a \\
\\
\frac{J_c}{\pi} \arctanh \frac{c |y|}{\sqrt{ \left(y^2 - b^2 \right)}} & y>a
\end{cases}
\end{align}
$J_c$ stands for the critical sheet current in units of $\unit{A/m}$ and the parameters $b$ and $c$ can be represented as
\begin{align*}	
b &= a/\cosh{\frac{\pi H_a}{J_c}} \\
c &= \tanh{\frac{\pi H_a}{J_c}},
\end{align*}
where $a$ corresponds to the halfed sample length.
The parameter $b$ can be interpreted as a lateral penetration depth of the externally applied magnetic field $H_a$, indicating how far $H_a$ penetrates from the sides into the sample. 
We fit our experimental data using this model in order to quantify the critical current density $j_c = J_c/d$, in which the flux lines start to move under the action of Lorentz force.
Note that, similar to the observation in Fig.~\ref{abb: main3}, the magnetic field is not vanishing inside the superconductor.
Therefore, we restrict our fit within the limit where the magnetic field is reduced to $18\%$ of the maximum observed field.  
This model agrees very well with our data for both $y<a$ (Fig.~\ref{abb: main4}(b), left panel) and for $y>a$ (Fig.~\ref{abb: main4}(b), right panel). 
The corresponding fitting paramaters for fixed values of $H_a = \unit[4.2]{mT}$ and $a=\unit[5000]{\mu m}$ are, $b = \unit[4709 \pm 2.2]{\mu m}$, $c = 0.335 \pm 0.0059$ and $J_c = \unit[27997 \pm 2249]{A/m}$.
Knowing $J_c$ and the sample thickness $d = \unit[20]{nm}$ we have extracted the critical current density as $j_c = \unit[1.4 \cdot 10^{8}]{A/cm^2}$. This value is in very good agreement with recently reported values for LSCO nanowires \cite{Lit} indicating the NV fluorescence drop as reliable quantity for characterizing thin film superconductors.
Also note, that the corresponding $j_c$ value agrees very well with a calculation based on our SQUID measurement in Fig.~\ref{abb: main2}.
The magnetic moment $m$ can be obtained using Brandt's analytical model and the corresponding $J_c$ with \cite{Brandt}
\begin{align}
m = J_c a^3 \tanh{\frac{\pi H_a}{J_c}}.
\end{align}
resulting into a magnetic moment of $m\approx \unit[0.012]{emu}$.
This is comparable with our SQUID measurement in Fig.~\ref{abb: main2}(b), in which the magnetic moment ranges from $\unit[0.0075]{emu}$ to $\unit[0.0052]{emu}$.
The slight mismatch in $m$ can be explained by the performed calibration measurement. The PL drop to magnetic field strength conversion in Fig.~\ref{abb: main1}(d) assumes a magnetic field applied in z-direction ($B_z$). This is a fair assumption for the region $y<a$ and $y>a$ (far inside and outside of the superconductor). However, in close proximity of the LSCO edge ($y=a$) the magnetic field is forced to curl around the edge of the superconductor. Therefore, it is expected, that the off-axis component of the magnetic field is not pointing only in the z-direction. Instead, a strong in-plane component has to be assumed. This fact can explain the overestimation of $j_c$.     
\section{Conclusion}
In conclusion, we have presented a microwave-free NV center based method for characterizing relative magnetic field changes. In order to demonstrate this, we investigated a superconducting LSCO thin film. So far, the NV center has mainly been used as magnetic field sensor in a resonant measurement scheme utilizing microwave excitations. Here we extend its application by employing the magnetic field dependent PL of an NV center ensemble in a microwave-free manner providing a direct manifestation of the Meissner screening in our LSCO sample. 
To the best of our knowledge, this is the first demonstration of NV center magnetometry on a superconducting sample that does not require resonant microwave pulse schemes, apart from calibrating the absolute magnetic field. The PL calibration can either over-or underestimate the magnetic field strength to a certain extent. However, this is not important for measurements relying on relative strength changes of the magnetic field such as in the case of the Meissner state in superconductors. The presented method enables a fast and precise measurement of such relative magnetic changes excluding heating effects on the sample. Furthermore, the magnetic field dependent NV emission used in this work can be extended further by combining optical pump probe spectroscopy, thereby enabling access to dynamical systems with fast timescales. With the NV center as nanoscale magnetic field sensor, this opens up a promising avenue for exploring a multitude of problems in condensed-matter physics such as non-equilibrium collective phenomena including the vortex formation and motion in type II superconductors.

\section*{Acknowledgments}
We acknowledge financial support for this work provided by the European Research Council (ERC) under the European Union’s Horizon 2020 research and innovation programme (grant agreement No. 742610), EU via project ASTERIQS, Max Planck Society and DFG via FOR 2724.
\section*{Data Availability}
The data that support the findings of this study are available from the corresponding author upon reasonable request. 
\section*{References}
\bibliography{aipsamp}

\end{document}